\documentclass[letter,aps,showpacs,twocolumn,floatfix,amsmath,amssymb,nolinenumbers,nofootinbib]{revtex4-1}
\usepackage{amssymb}
\usepackage{amsmath}
\usepackage{epsfig}
\usepackage{graphicx}
\usepackage{latexsym}
\usepackage{bm,latexsym}
\usepackage{mathrsfs}

\def\Journal#1#2#3#4{{#1} {\bf #2}, #3 (#4)}

\def\CQG{{Class. Quant. Grav.}}

\def\LivRev{{Living Rev. Relativity}} 
\def\PLB{{Phys. Lett.}  B}

\def\PRL{Phys. Rev. Lett.}

\def\PRD{{Phys. Rev.} D}

\def\RMP{{Rev. Mod. Phys.}}

\begin{document}


\title{Robust approach to $f(R)$ gravity}

\author{Luisa G. Jaime$^{1,2}$}
\email{lgjaime@astro.unam.mx}

\author{Leonardo Pati\~no$^2$}
\email{leopj@ciencias.unam.mx}

\author{Marcelo Salgado$^1$}
\email{marcelo@nucleares.unam.mx}

\affiliation{$^1$Instituto de Ciencias Nucleares, Universidad Nacional
Aut\'onoma de M\'exico, A.P. 70-543, M\'exico D.F. 04510, M\'exico \\
$^2$ Facultad de Ciencias, Universidad Nacional
Aut\'onoma de M\'exico, A.P. 50-542, M\'exico D.F. 04510, M\'exico }


\date{\today}

    
\begin{abstract}
We consider metric $f(R)$ theories of gravity without mapping them to their scalar-tensor counterpart, but using the 
Ricci scalar itself as an ``extra'' degree of freedom. This approach avoids then the introduction of 
a scalar-field potential that might be ill defined (not single valued). In order to explicitly show the usefulness 
of this method, we focus on static and spherically symmetric spacetimes and deal with the recent controversy about 
the existence of extended relativistic objects in certain class of $f(R)$ models.
\end{abstract}




\maketitle


\section{Introduction}
\label{sec:introduction}
Modified $f(R)$ theories of gravity are metric theories which postulate an {\it a priori} arbitrary function of the Ricci scalar 
$R$ as the Lagrangian density. These theories have been proposed to explain the late time accelerated expansion of the Universe 
as well as a mechanism to produce inflation in the early Universe in terms of geometry instead of introducing any dark energy 
ideas or a scalar inflaton~\cite{accexp}. Thanks to the above properties, these theories have become one of the most popular 
alternative theories of gravity. In recent years, several specific $f(R)$ models have been analyzed in different settings~
(see Ref.~\cite{f(R)}, for a review). Although the early models were ruled out for failing several tests (like the Solar System 
experiments) new proposals were put forward to overcome such drawbacks. Nevertheless, a considerable amount of analysis and 
observational confrontation is still required in order to compare the preliminary successes of such theories with the great 
achievements of general relativity (GR). In particular, the models that were claimed to pass several 
cosmological and local tests have been scrutinized in the strong gravity regime only recently. 
In fact, the first studies concerning their ability to describe relativistic extended objects, like neutron stars, 
seem to give contradictory results. Using a $f(R)$ model proposed by 
Starobinsky~\cite{Starobinsky2007}, Kobayashi and Maeda~\cite{Kobayashi2008} found that such relativistic objects were 
difficult to construct since a {\it curvature singularity} developed within the object. Later, this issue was reanalyzed 
by several authors ~\cite{Babichev2009,Upadhye2009}, who found that singularity-free relativistic objects can indeed be 
constructed, arguing that the conclusion reached in ~\cite{Kobayashi2008} was due to a bad choice of the matter model 
({\it i. e.} the equation of state, hereafter EOS)~\cite{Babichev2009}, while in ~\cite{Upadhye2009} it was claimed that 
a ``chameleon'' is the responsible for the existence of such ``stars,'' regardless of the EOS.

A common feature to all of the aforementioned works~\cite{Kobayashi2008,Babichev2009,Upadhye2009} is that the analysis 
was performed by mapping the Starobinsky model to a scalar-tensor theory of gravity (STT)~({\it cf.} \cite{Faraoni2007}). 
Under this mapping, the Ricci scalar has a behavior of the sort $R\sim 1/(\chi - \chi_0)$, where $\chi:= \partial_R f$ 
is the scalar-field associated with the corresponding STT and $\chi_0={\rm const.}$ The key point is to determine if the 
dynamics of $\chi$ leads it or not to the value $\chi=\chi_0$ within the spacetime generated by the relativistic object. 
Irrespective of the different results and confusing explanations obtained in ~\cite{Kobayashi2008,Babichev2009,Upadhye2009}, 
we will argue that their conclusions are rather questionable due to the fact that the mapping to the STT is ill defined. 
To be more specific, the scalar-field potential used to study the dynamics of $\chi$ is not single valued and possesses 
pathological features. Since similar kind of singularities were also found in the cosmological setting~\cite{Frolov2008}, 
it is then worrisome that the ill-defined potential play such a crucial role in those analyses ({\it cf.} 
Refs.~\cite{Miranda2009,Goheer2009} for further criticisms. See also ~\cite{Nojiri2008} and references therein for 
a more detailed discussion on cosmological singularities in $f(R)$ theories).

The goal of this communication is threefold: 1) to propose a straightforward and robust approach which consists in recasting the 
field equations in a more suitable way without mapping the original $f(R)$ theory to any scalar-tensor counterpart. 
This method dispenses us from dealing with ill-defined quantities that might arise when performing such transformation; 
2) to reanalyze carefully the issue about the existence of relativistic extended objects using our approach; 3) to 
stress that the analysis of $f(R)$ theories based on the STT analogue 
with ill-defined quantities is not trustworthy, and that in cases where the STT approach is well defined 
(this depends on the specific $f(R)$ model) it turns out to be rather convoluted and not very insightful. 
Since $f(R)$ theories are still under close examination, a sound approach is 
required to treat them appropriately. This is the first step in that direction.


\section{$f(R)$ theories of gravity}
\label{sec:ADM}
The general action for a $f(R)$ theory of gravity is given by
\begin{equation}
\label{f(R)}
S[g_{ab},{\mbox{\boldmath{$\psi$}}}] =
\!\! \int \!\! \frac{f(R)}{2\kappa} \sqrt{-g} \: d^4 x 
+ S_{\rm matt}[g_{ab}, {\mbox{\boldmath{$\psi$}}}] \; ,
\end{equation}
where $R$ stands for the Ricci scalar, $\kappa:= 8\pi G_0$, and 
${\mbox{\boldmath{$\psi$}}}$ represents collectively the matter
fields (we use units where $c=1$). The field equation arising from varying the action Eq.~(\ref{f(R)}) with respect to the metric is
\begin{equation}
\label{fieldeq1}
f_R R_{ab} -\frac{1}{2}fg_{ab} - 
\left(\nabla_a \nabla_b - g_{ab}\Box\right)f_R= \kappa T_{ab} \; ,
\end{equation}
where $f_R:=\partial_R f$ and $\Box= g^{ab}\nabla_a \nabla_b$. It is straightforward to 
write the above equation in the following way
\begin{eqnarray}
\label{fieldeq2}
&& \!\!\!\!\!\!\!\!\!\!
 f_R G_{ab} - f_{RR} \nabla_a \nabla_b R - 
 f_{RRR} (\nabla_aR)(\nabla_b R) \nonumber \\ 
&& \!\!\!\!\!\!\!\!\!\!
 + g_{ab}\left[\frac{1}{2}\left(Rf_R- f\right)
+ f_{RR} \Box R + f_{RRR} (\nabla R)^2\right] = \kappa T_{ab} \; ,
\end{eqnarray}
where $(\nabla R)^2:= g^{ab}(\nabla_aR)(\nabla_b R)$. 
Taking the trace of this equation yields
\begin{equation}
\label{traceR}
\Box R= \frac{1}{3 f_{RR}}\Big{[}
\kappa T - 3 f_{RRR} (\nabla R)^2 + 2f- Rf_R
\Big{]} \; ,
\end{equation}
where $T:= T^a_{\,\,a}$. Finally, using Eq.~(\ref{traceR}) in ~(\ref{fieldeq2}), we find
\begin{eqnarray}
\label{fieldeq3}
&& G_{ab} = 
\frac{1}{f_R}\Bigl{[} f_{RR} \nabla_a \nabla_b R +
 f_{RRR} (\nabla_aR)(\nabla_b R) \nonumber\\ 
&&  - \frac{g_{ab}}{6}\Big{(} Rf_R+ f + 2\kappa T \Big{)} 
+ \kappa T_{ab} \Bigl{]} \; .
\end{eqnarray}
Equations~(\ref{traceR}) and ~(\ref{fieldeq3}) are the basic equations for $f(R)$ 
gravity that we propose to treat in every application, instead of transforming them to 
STT. Now, several important remarks are in order regarding this set of equations. 
First, aside from the case $f(R)=R$, where the field equations reduce to those of GR, 
one is to consider functions $f(R)$ such that $f_R,f_{RR} > 0$ 
({\it i. e.} monotonically growing and convex $f(R)$ functions) in order to avoid potential blowups 
in the field equations. These two conditions have been considered previously; the first one leads to a
a positive definite gravitational ``constant'' $G_{eff}= G_0/f_R$ when regarding $f(R)$ theories as effective 
theories. The condition $f_{RR}>0$ was suggested to avoid gravitational instabilities~\cite{Dolgov2003}. 
Second, Eq.~(\ref{fieldeq3}) supplies a second order equation for the metric provided that the Ricci 
scalar is considered as an independent field. This is possible thanks to Eq.~(\ref{traceR}) which 
provides the information needed to solve Eq.~(\ref{fieldeq3}) for the metric alone (given a matter source).

As will become evident below, this approach is rather clean and free of the 
pathologies that can arise when mapping $f(R)$ gravity to STT.


\section{Static and spherically symmetric spacetimes}
\label{sec:numres}
In order to give some insight to the usefulness of our approach, we consider 
a static and spherically symmetric
(SSS) spacetime with a metric given by 
$ds^2 = - n(r) dt^2  + m(r) dr^2+ r^2 \left(d\theta^2 + \sin^2\theta d\varphi^2\right)$, 
where the metric coefficients $n$ and $m$ are functions of the
coordinate $r$ solely. The Ricci scalar as well as the matter variables will also be 
functions of $r$. The analysis of this kind of spacetimes is interesting in many ways; however, we will focus here 
only on the issue about the existence (or absence thereof) of well behaved compact objects. Equation~(\ref{traceR}) yields
\begin{eqnarray}
\label{traceRr}
R'' &=& \frac{1}{3f_{RR}}\Big{[}m(\kappa T+2f- Rf_{R}) - 3f_{RRR}R'^2\Big{]} \nonumber \\
&& +\left(\frac{m'}{2m}-\frac{n'}{2n}-\frac{2}{r}\right)R' \;.
\end{eqnarray}
(where $^\prime:= d/dr$). From the $t-t$, $r-r$, and $\theta-\theta$ components of Eq.~(\ref{fieldeq3}) and using 
also Eq.~(\ref{traceRr}) we find
\begin{eqnarray}
\label{mprime}
m' &=& \frac{m}{r(2f_{R}+rR'f_{RR})} \Biggl{\{} 2f_{R}(1-m)-2mr^2 \kappa T^{t}_{\,\,t} 
 \nonumber \\
&&\!\!\!\!\!\!\!\!\!\!\!\!\!\!\!\!\! +\frac{mr^2}{3}(Rf_{R}+f+2\kappa T) 
+ \frac{rR'f_{RR}}{f_{R}}\Bigl{[}\frac{mr^2}{3}(2Rf_{R}-f+\kappa T) \nonumber \\
&&\!\!\!\!\!\!\!\!\!\!\!\!\!\!\!\!\!
-\kappa mr^2(T^{t}_{\,\,t}+T^{r}_{\,\,r})+2(1-m)f_{R}+2rR'f_{RR}\Bigr{]} \Biggr{\}} \; , 
\end{eqnarray}
\begin{eqnarray}
\label{nprime}
 n' &=& \frac{n}{r(2f_{R}+rR'f_{RR})} \Bigl{[} mr^2(f-Rf_{R}+2\kappa T^{r}_{\,\,r}) \nonumber \\
&& +2f_{R}(m-1)-4rR'f_{RR}  \Bigr{]} \; ,\\
\label{nbiprime}
n'' &=& \frac{2nm}{f_{R}} \Bigl{[}  \kappa T^{\theta}_{\,\,\theta}-\frac{1}{6}(Rf_{R}+f+2\kappa T)
+ \frac{R'}{rm}f_{RR}\Bigr{]} \nonumber \\
&& + \frac{n}{2r}\Bigl{[}2\left(\frac{m'}{m}-\frac{n'}{n}\right)+\frac{rn'}{n}\left(\frac{m'}{m}
+\frac{n'}{n}\right)\Bigr{]} \; .
\end{eqnarray}
Note that Eqs.~(\ref{nprime}) and (\ref{nbiprime}) are not independent. In fact, one has the freedom 
of using one or the other (see Sec.~\ref{sec:numerical results} below). Now, from the usual expression of $R$ 
in terms of the Christoffel symbols one obtains
\begin{eqnarray}
\label{Rr}
 R &=& \frac{1}{2r^2n^2m^2}\Bigl{[}4n^2m(m-1)+rnm'(4n+rn')  \nonumber \\
&& -2rnm(2n'+rn'')+r^2mn'^2\Bigr{]} \; .
 \end{eqnarray}
As one can check by a direct calculation, using Eqs.~(\ref{mprime})$-$(\ref{nbiprime}) in 
Eq.~(\ref{Rr}) leads to an identity $R\equiv R$. This result confirms the consistency of our equations. 
When defining the first order variables $Q_n= n'$ and $Q_R:= R'$, Eqs.~(\ref{traceRr})$-$(\ref{nbiprime})
have the form $dy^i/dr= {\cal F}^i(r,y^i)$, where $y^i= (m,n,Q_n,R,Q_R)$, and therefore can be solved numerically. 
As far as we are aware, the system 
~(\ref{traceRr})$-$(\ref{nbiprime}) has not been considered previously (see Ref.~\cite{Nzioki2010} for an alternative 
approach). These equations can be used to tackle several aspects of SSS spacetimes in $f(R)$ gravity. Finally, we observe 
that for $f(R)=R$ the above equations also reduce to the well known equations of GR for SSS spacetimes. 
When dealing with extended objects, notably those described by perfect 
fluids, the above equations are supplemented by the matter conservation equations. The Bianchi identities imply the conservation 
of the effective energy-momentum tensor [which corresponds to the right-hand-side (rhs) of Eq.~(\ref{fieldeq3})] which 
together with Eqs.~(\ref{traceR}) and (\ref{fieldeq3}) lead to the equation of conservation $\nabla^a T_{ab}= 0$ 
for the matter alone. So for a perfect fluid with $T_{ab}= (\rho + p)u_a u_b + g_{ab} p$, the conservation equation 
leads to $p'= -(\rho + p) n'/2n$ [where $n'$ is given explicitly by the rhs of Eq.~(\ref{nprime})]. This 
is the modified Tolman-Oppenheimer-Volkoff equation of hydrostatic equilibrium to be complemented by an EOS.

In order to solve the differential equations some boundary conditions must be supplied, which in this case are rather 
regularity and asymptotic conditions. Regularity (smoothness) at $r=0$ implies 
the following expansion near $r=0$: $\phi(r)= \phi_0 + \phi_1 r^2/2 + {\cal O}(r^4)$ (where $\phi$ stands for $m,n,$ 
or $R$). This implies $m'=0=n'=R'$ at $r=0$. We choose $m(0)=1$ (local-flatness condition) and $n(0)=1$. 
The coefficients $\phi_0$ and $\phi_1$ associated with $m,n,R,$ and the matter variables (and which correspond to 
the values of these quantities and its second derivatives evaluated at $r=0,$ respectively) are related to each other. 
For instance, from the above power expansion and from Eqs.~(\ref{nbiprime}), (\ref{traceRr}) and (\ref{Rr}), 
we find $n''(0)=\frac{f^{0}-2f_{R}^{0}R^{0}-4kT^{0}+18k{T_{\theta}^{\theta}}^{0}}{9f_{R}^{0}}$ and 
$R''(0)= \frac{2f^{0}-f_{R}^{0}R^{0}+kT^{0}}{9f_{RR}^{0}}$, where the quantities at the rhs are evaluated at $r=0$
~\footnote{Note that in GR the field equations imply the algebraic relationship $R=-kT$, which determine the 
value of $R$ in terms of the matter content solely. In $f(R)$ theories this relationship is {\it differential} [{\it cf.} 
Eqs.~(\ref{traceR}) and (\ref{traceRr})], and therefore the conditions for $R(0)$, $R''(0)$ and even for 
$n''(0)$ are not fixed in advance by the matter content only. We can recover the usual expressions of GR when taking 
$f(R)=R$.}.
 
Now, as concerns the asymptotic conditions, we are interested in finding solutions that are asymptotically de Sitter, since 
the solutions are supposed to match a cosmological solution that represents the observed Universe. 
Therefore, we demand that asymptotically $R \rightarrow R_1$, where $R_1$ is a critical point 
(maximum or minimum) of a potential which is defined 
below. The value $R_1$ allows to define the effective cosmological constant as $\Lambda_{\rm eff} =R_1/4$ 
(like in GR with $\Lambda$)\footnote{One can easily show that the system Eqs.~(\ref{traceRr})$-$(\ref{nbiprime}) in 
vacuum has the exact de Sitter solution $n(r)=m(r)^{-1}=1 - \Lambda_{\rm eff} r^2/3$, $R=R_1={\rm const.}$ with 
$\Lambda_{\rm eff} =R_1/4$ and $R_1= 2f(R_1)/f_R(R_1)$, which corresponds precisely where $dV(R)/dR=0$, 
as defined in the main text.}. We mention that at the end of the numerical integration, we 
rescale $n(r)$ in order to get the canonical asymptotic behavior $n(r)\sim 1 - \Lambda_{\rm eff} r^2/3$. This rescaling amounts 
to redefine the $t$ coordinate. The asymptotic behavior $R \rightarrow R_1$ depends on the value of $R(0)$.
The details of the correlation between $R(0)$ and $R_1$ depend on the matter model, and
once this latter is fixed  (for instance, an incompressible fluid with a given central pressure) the value $R(0)$ can 
be found by a shooting method~\cite{Numrec}. Since we are interested in finding an exterior solution of Eq.~(\ref{traceRr}) with 
$R \rightarrow R_1$ (and $R'\rightarrow 0$) asymptotically, we observe that such a solution exists if $R_1$ corresponds to a 
critical point of the ``potential'' $V(R)= -R f(R)/3 + \int^R f(x) dx$. That is, $R_1$ is a point where 
$dV(R)/dR =\left(2f-f_{R}R\right)/3$ vanishes [assuming $f_{RR}(R_1)\neq 0$ in Eq.~(\ref{traceRr})]
. This potential is radically different from the scalar-field potential that arises under the STT map. 
Furthermore, $V(R)$ is as well defined as the function $f(R)$ itself.

A technical difficulty that one faces when integrating the equations for neutron star models that asymptotically match 
realistic values of the cosmological constant is that {\it a fortiori} two completely different scales are involved. 
On one hand $\rho(0)\sim \rho_{\rm nuc} \sim 10^{14} {\rm g\,cm^{-3}}$; on the other 
$\tilde\Lambda \sim 10^{-29} {\rm g\,cm^{-3}}$ ($\tilde \Lambda= \Lambda/G_0$). That is, there are around 43 orders of magnitude 
between the typical density within a neutron star and the average density of the Universe. This ratio between densities 
naturally appears in the equations since the parameters which define the specific $f(R)$ 
theory are of the order of $\tilde\Lambda$, while the appropriate dimension within neutron stars is $\rho_{\rm nuc}$. So 
in units of $\rho_{\rm nuc}$, the cosmological constant turns out to be ridiculously small, while in units of $\tilde\Lambda$, 
$\rho(r)$ and $p(r)$ turn out to be ridiculously large within the neutron star. So far, the authors that have studied 
relativistic objects in $f(R)$ gravity~\cite{Kobayashi2008,Babichev2009,Upadhye2009,Miranda2009} have faced 
this kind of technical problem, and, in order to avoid it, they have only constructed objects which are far from 
representing neutron stars embedded in a realistic de Sitter background. This means that either the background is 
realistic while the objects are ridiculously large (several orders of magnitude larger than a real neutron star) and 
not very dense, or the objects are realistic but $\tilde\Lambda_{\rm eff}$ is far too large. 
Such objects are relativistic in the sense that their pressure is of the same order 
of magnitude as their energy-density, and the ratio between a suitably defined mass and radius 
$G_0 {\cal M}/{\cal R}$ is not far from $4/9$.


\section{Numerical Results}
\label{sec:numerical results}
\begin{figure}[h!]
\includegraphics[width=0.47\textwidth,height=40mm]{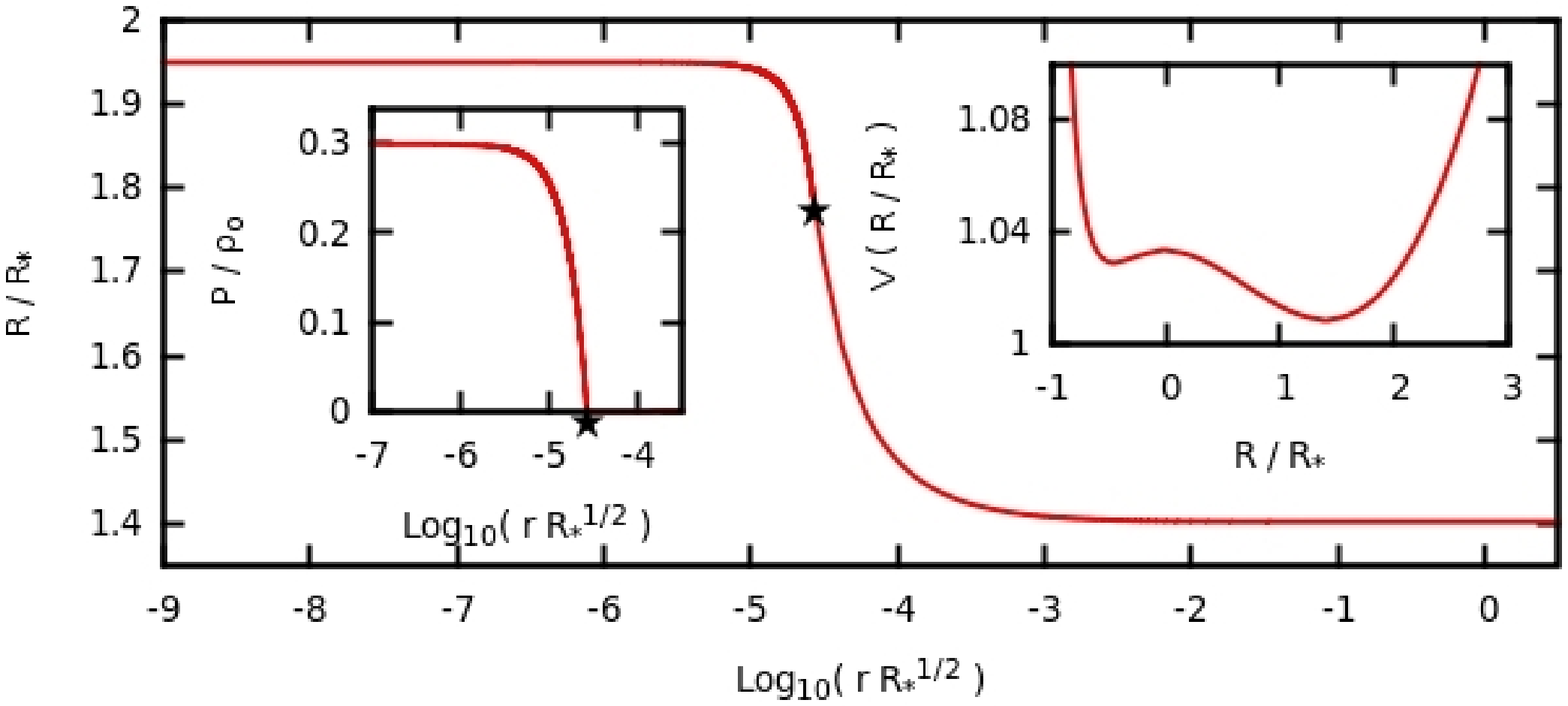} 
\includegraphics[width=0.47\textwidth,height=40mm]{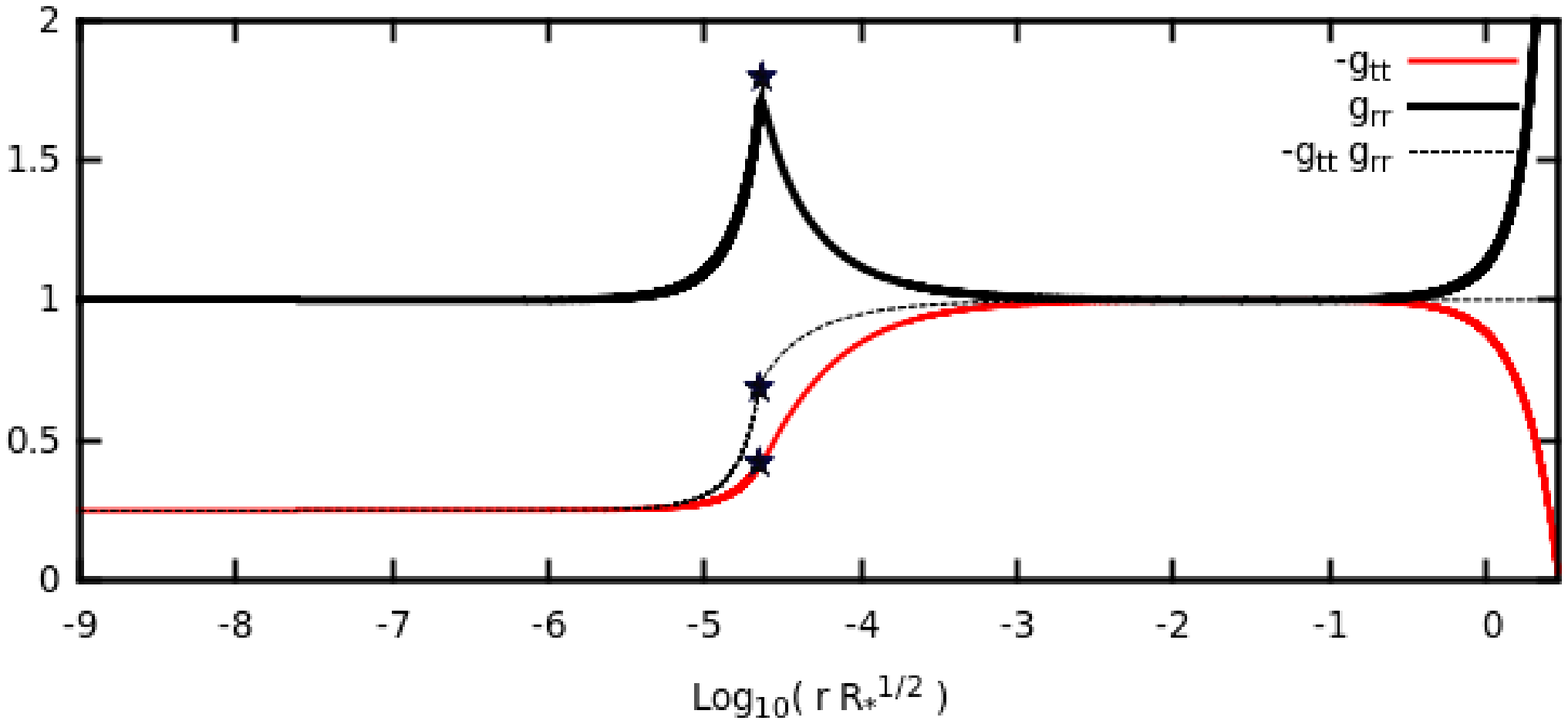}
\caption{(Top) Ricci scalar and pressure (left inset) as functions of $r$ taking $\rho=\rho_0={\rm const.}=5\times10^7 R_1/G_0$, 
$p(0)=0.3 \rho_0$, and $R_1/R_*\sim 1.405$ as in \cite{Miranda2009}. The stars depict the 
location of the object's surface. Potential $V(\tilde R)$ (right inset) in units of $R_*^2$, for the model 
$f(R)=R-\alpha R_*{\rm ln}\left(1+ R/R_*\right)$ with $\alpha= 1.2$. (Bottom) metric potentials associated with the 
solution for $R$.}
\label{fig:sols1}
\end{figure}

\begin{figure}[h!]
\includegraphics[width=0.47\textwidth,height=40mm]{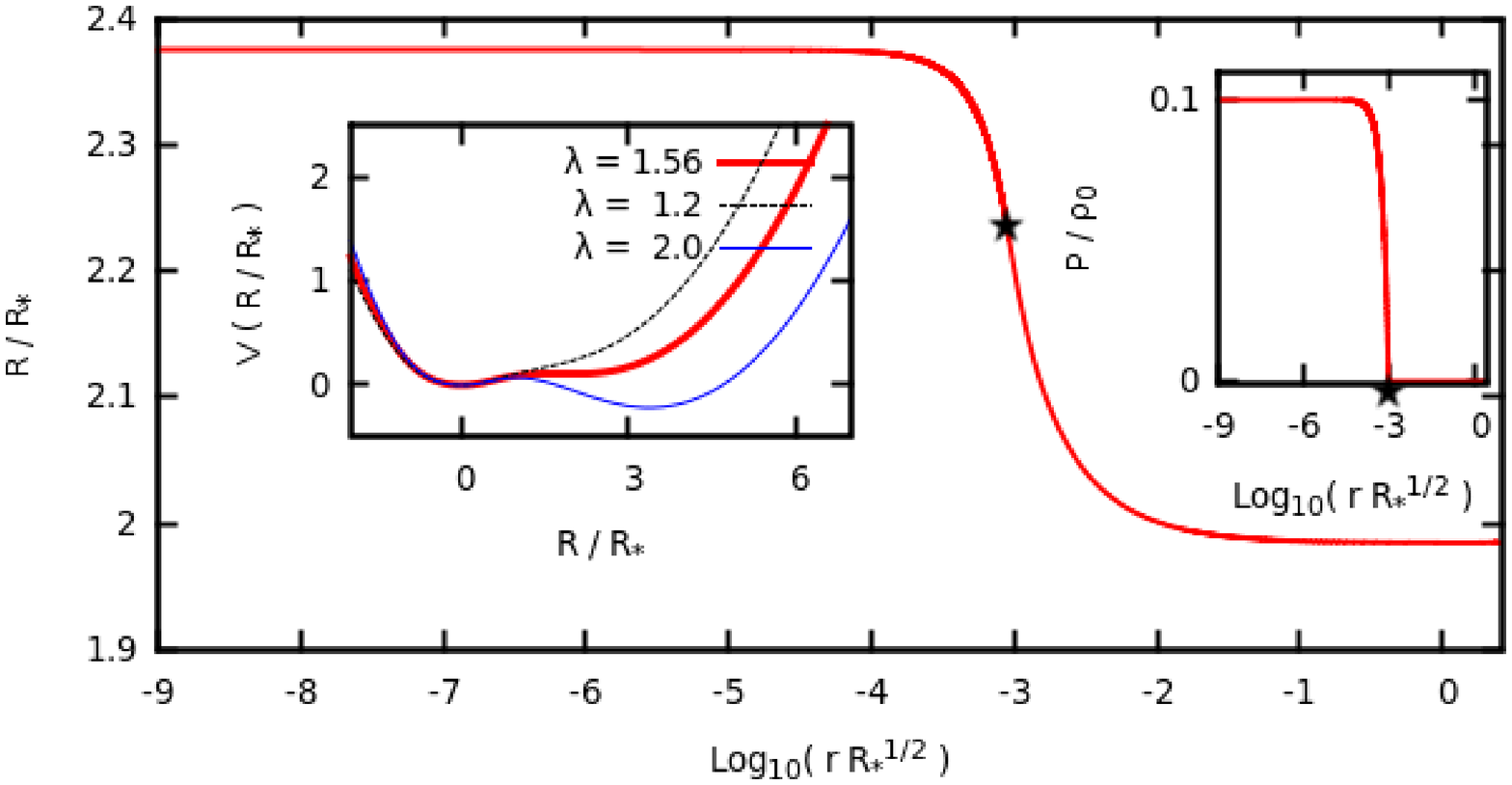} 
\includegraphics[width=0.47\textwidth,height=40mm]{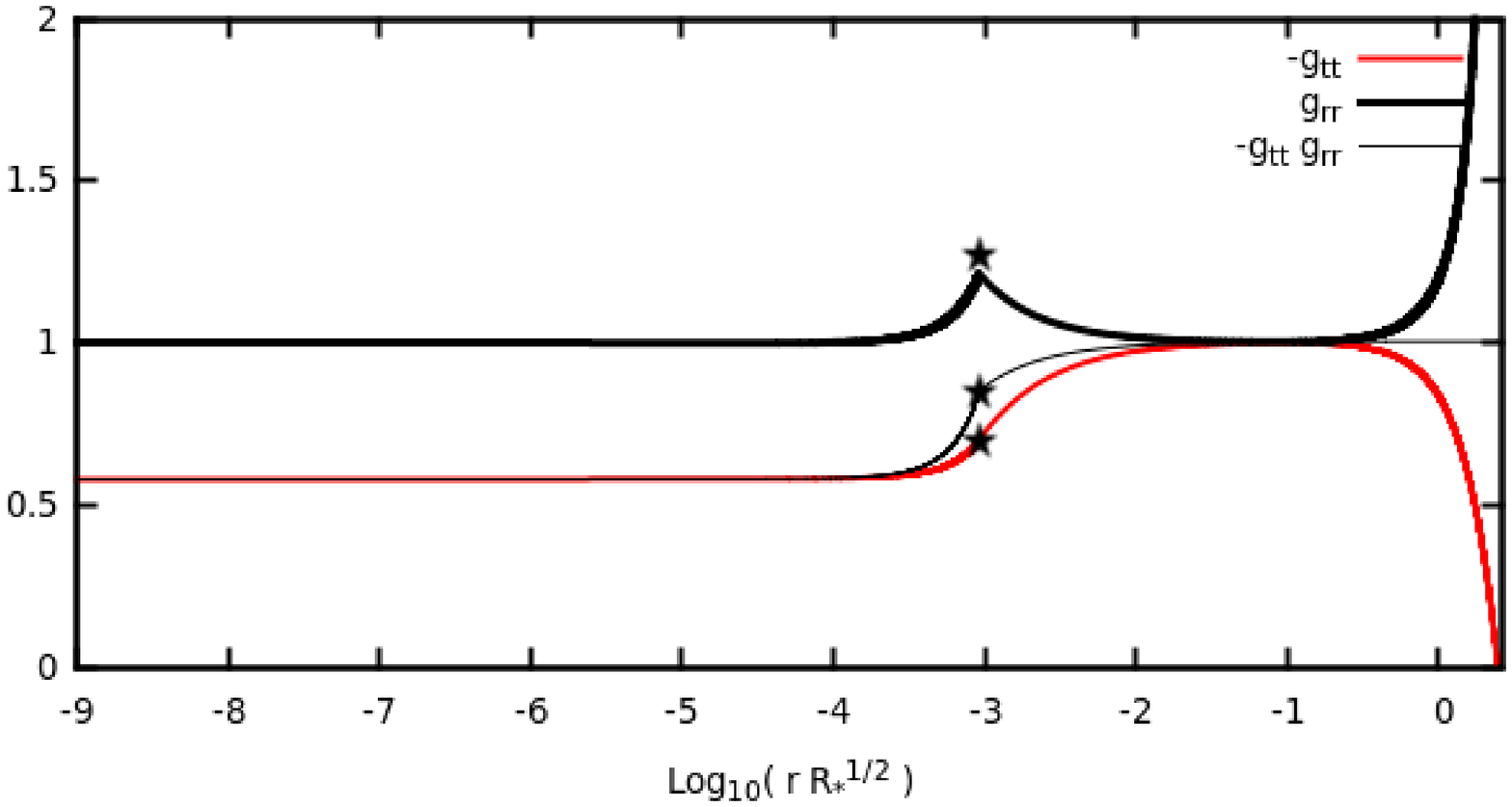}
\caption{Similar as Fig.~\ref{fig:sols1} for the Starobinsky model, with $\lambda=1.56$ 
and $\beta=1$  taking $\rho_0=10^6 R_1/(16\pi G_0) $, $p(0)=0.1 \rho_0$ with $R_1/R_*\sim 1.983$.}
\label{fig:sols2}
\end{figure}

In order to test our method, we used first $f(R)=R-\alpha R_*{\rm ln}\left(1+ R/R_*\right)$ ($\alpha,R_*$ are positive 
constants; $R_*$ sets the scale), which was proposed in Ref.~\cite{Miranda2009}. In that analysis the authors mapped 
the theory to the STT counterpart. However, unlike the Starobinsky model (see below), in this case the resulting 
scalar-field potential turns to be single valued and the authors of \cite{Miranda2009} did not find any singularity 
within the object. Under our approach, we associate to this $f(R)$ the potential 
$V(\tilde R) =\frac{R_*^2}{6}\Big{\{} (1+\tilde R)(\tilde R + 6\alpha-1) 
 -2\alpha(3+2\tilde R){\rm ln}(1+ \tilde R) \Big{\}}$, where $\tilde R= R/R_*$.
This potential has several critical points. Figure~\ref{fig:sols1} (top panel, right inset) depicts the potential for $\alpha=1.2$. 
Equations~(\ref{traceRr})$-$(\ref{nprime}) were integrated using a four-order Runge-Kutta algorithm by assuming an 
incompressible fluid with the same density and central pressure as in~\cite{Miranda2009}. By implementing a shooting 
method~\cite{Numrec}, we found a solution for $R$ that goes from the global minimum $V(\tilde R_1)$ of the potential to a positive value. 
The minimum at $\tilde R_1$ corresponds to the de Sitter value and gives rise to the effective cosmological constant 
$\Lambda_{\rm eff}= R_1/4$. Figure~\ref{fig:sols1} (top panel) depicts the solution for $R$, the pressure (top panel, left inset) 
and the metric potentials (bottom panel) plotted up to the value where $n(r)$ reaches the cosmological horizon given by $n(r_h)=0$. 
We checked that asymptotically the metric potentials matched perfectly well the de Sitter solution 
$n(r)=m(r)^{-1}=1 - \Lambda_{\rm eff} r^2/3$. We also verified that our solution corresponds to the one found 
in~\cite{Miranda2009} and also confirmed that no singularity 
whatsoever was encountered. Furthermore, we checked that the numerical solutions found using the system 
Eqs.~(\ref{traceRr})$-$(\ref{nprime}) and the system Eqs.~(\ref{traceRr}),(\ref{mprime}) and (\ref{nbiprime}) 
gave the same results within a relative error no larger than $\sim 10^{-6}$ and that for both systems the identity 
Eq.~(\ref{Rr}) was satisfied within a relative error of $\sim 10^{-10}$. In this way, we ensured that no mistakes were 
introduced in the code during the typing of the equations. We emphasize that for this model the condition $f_{RR}>0$ 
is satisfied by construction but the condition $f_R>0$ is not in general. However, for the solutions we found, this latter is always 
satisfied, particularly at $R_1$. Having analyzed the solutions for this $f(R)$, we turned our 
attention to the Starobinsky model given by $f(R)= R - \lambda R_*\left\{1-\left[1+ (R/R_*)^2\right]^{-\beta}\right\}$ 
with $\beta=1$. This is the model analyzed in~\cite{Kobayashi2008,Babichev2009,Upadhye2009} using the 
transformation to STT. For such a model, the associated scalar-field potential is multiple-valued 
({\it cf.} Fig. 1 in~\cite{Frolov2008} and Fig. 2 in~\cite{Kobayashi2008}). 
Despite the potential being ill-defined, the authors in \cite{Kobayashi2008} argue that in the particular region where the 
scalar-field is evaluated, the potential is nevertheless well defined. 
It seems suspicious that in~\cite{Babichev2009,Upadhye2009} the odd features of the potential are not mentioned at all. 
Even if one tried to ignore the region where the potential is pathological and 
consider only the single-valued region, there is no guarantee that the ill-defined region will not play any 
role in more general settings than the very specific case of static and spherically symmetric 
compact objects. Modifications of the Starobinsky model by adding a quadratic term do not cure this pathology, as one can see 
in Fig. 5 of~\cite{Dev2008}.  As we stressed before, 
the solutions found in ~\cite{Kobayashi2008,Babichev2009,Upadhye2009} gave 
rise to a controversy since in~\cite{Kobayashi2008} the authors found a singularity in the Ricci scalar within the extended 
object, while in~\cite{Babichev2009,Upadhye2009} the authors did not. Following our approach, we get 
$V(\tilde R) =\frac{R_*^2}{3} \left\{\rule{0mm}{0.5cm}
\frac{\tilde R}{2}\left[\tilde R-4\lambda - 2\lambda \left(1+{\tilde R}^2\right)^{-1} \right]
+3\lambda\,{\rm arctan}(\tilde R) \right\}$. 
This potential has a rich structure depending on the value of $\lambda$ (such structure arises also for the potential discussed above 
for different values of $\alpha$). Figure~\ref{fig:sols2} (top panel, left inset) depicts the potential for three values of $\lambda$. 
There is a value $\lambda_{\rm crit}= 8/\sqrt{27}$ below which the potential has only one critical point 
(the global minimum) located at $R=0$. In this regime ($\lambda < \lambda_{\rm crit}$), we have found only oscillatory solutions around 
the global minimum. For $\lambda > \lambda_{\rm crit}$, three critical points appear, 
with a minimum always located at $R=0$ ($R_0$). In view of this structure, several kinds of solutions are possible for any given 
value of $\lambda > \lambda_{\rm crit}$. We have found solutions where $R$ asymptotically approaches a minimum 
of $V(R)$ at $R_1\neq 0$, the local maximum, and also solutions where $R$ oscillates asymptotically around $R_0$. 
An asymptotic oscillatory behavior around $R_0$ seems {\it a priori} not adequate to produce 
the required de Sitter background; we plan to analyze those solutions in more detail elsewhere. Here, we are only interested in 
showing a solution for which $R$ goes to $R_1$ (a local minimum). Figure~\ref{fig:sols2} shows the numerical solution which is 
asymptotically de Sitter. As one can appreciate, no singularity in the Ricci scalar was encountered in this solution 
and $R$ never crosses the two real-valued zeros of $f_{RR}$ (which correspond to $R= \pm R_*/\sqrt{3}$) 
nor those of $f_R$ at which blowups in the equations can be produced.  Even if for the static and spherically symmetric solutions, 
one can avoid the zeros of $f_{R}$ and $f_{RR}$, the Starobinsky model does not satisfy the conditions $f_R,f_{RR}>0$ in general, 
and so, in other settings ({\it e. g.} cosmology) one has to keep this in mind when solving the equations. 

\section{Conclusions}
\label{sec:discussion}
We have devised a straightforward and robust approach to treat $f(R)$ metric gravity without resorting to the 
usual mapping to scalar-tensor theories. With this method, it is possible to analyze in a rather transparent and well defined way 
several aspects of these alternative theories. In particular, we focused on the existence of relativistic extended objects 
embedded in a de Sitter background and concluded that for some $f(R)$ models such objects can be constructed without ambiguity 
and without resorting to any dubious explanations based on the use of ill-defined quantities.
It seems that the analysis of the solutions presented here as well as other ({\it i. e.} solutions that admit different de 
Sitter backgrounds, even possibly Minkowski backgrounds) can only be carried out easily following our method 
since our potential $V(R)$ depicts many features in a rather clear and clean fashion that allows us to identify the 
critical points which $R$ can reach asymptotically.

Building realistic neutron stars with a realistic de Sitter background in $f(R)$ gravity still remains a technical challenge. 
In the near future, we plan to study in more detail this issue along with other aspects of $f(R)$ gravity using our approach.

\acknowledgments
Special thanks are due to S. Jor\'as and D. Sudarsky. This work was partially supported by CONACyT Grant No. SEP-2004-C01-47209-F 
and by DGAPA-UNAM Grants No. IN119309-3, No. IN115310, and No. IN-108309-3.



\end{document}